\documentclass[apjl]{emulateapj}
\usepackage{apjfonts}
\usepackage{aas_macros}

\usepackage{bm}
\usepackage{amsmath}
\usepackage{rotating}

\newcommand\lsim{\mathrel{\rlap{\lower4pt\hbox{\hskip1pt$\sim$}}
        \raise1pt\hbox{$<$}}}
\newcommand\gsim{\mathrel{\rlap{\lower4pt\hbox{\hskip1pt$\sim$}}
        \raise1pt\hbox{$>$}}}
\newcommand\propsim{\mathrel{\rlap{\lower4pt\hbox{\hskip1pt$\sim$}}
        \raise1pt\hbox{$\propto$}}}
\newcommand{\Msun}{\mathrm{M}_{\odot}}

\newcommand{\A}{\mathcal{A}}
\newcommand{\dL}{d_{\rm L}}
\newcommand{\sech}{\mathrm{sech}}
\newcommand{\EF}{\mathrm{EF}}
\newcommand{\FH}{\mathrm{FH}}
\renewcommand{\c}{\mathrm{c}}
\newcommand{\G}{\mathrm{G}}
\newcommand{\Gpc}{\,\mathrm{Gpc}}
\newcommand{\Mpc}{\,\mathrm{Mpc}}
\newcommand{\Hz}{\,\mathrm{Hz}}

\begin{document}

\title{High Frequency Gravitational Waves from Supermassive Black Holes:
\\ Prospects for LIGO--Virgo Detections}

\author{Bence Kocsis}
\email{bkocsis@cfa.harvard.edu}
\affil{Harvard-Smithsonian Center for Astrophysics, 60 Garden Street, Cambridge, MA 02138, USA}
\submitted{Submitted to ApJ}
\date{\today}

\begin{abstract}
{
It is commonly assumed that ground-based gravitational wave (GW) instruments will not be sensitive to supermassive black holes (SMBHs) because the characteristic GW frequencies are far below the $\sim 10-1000$~Hz sensitivity bands of terrestrial detectors.  Here, however, we explore the possibility of SMBH gravitational waves to leak to higher frequencies. In particular, if the high frequency spectral tail asymptotes to $\tilde{h}(f)\propto f^{-\alpha}$, where $\alpha\leq 2$, then the spectral amplitude is a constant or increasing function of the mass $M$ at a fixed frequency $f\gg c^3/G M$.  This will happen if the time domain waveform or its derivative exhibits a discontinuity.} Ground based instruments could search for these universal spectral tails to detect or rule out such features irrespective of their origin. We identify the following processes which may generate high frequency signals: (i) gravitational bremsstrahlung of ultrarelativistic objects in the vicinity of a SMBH, (ii) ringdown modes excited by an external process that has a { high frequency component or} terminates abruptly, (iii) gravitational lensing echos and diffraction. We estimate the order of magnitude of the detection signal to noise ratio for each mechanism (i, ii, and iii) as a function of the waveform parameters. In particular for (iii), SMBHs produce GW echos of inspiraling stellar mass binaries in galactic nuclei with a delay of a few minutes to hours. The lensed primary signal and GW echo are both amplified if the binary is within a $\sim 10\,{\rm deg}\, (r/100M)^{-1/2}$ cone behind the SMBH relative to the line of sight at distance $r$ from the SMBH. For the rest of the binaries near SMBHs, the amplitude of the GW echo is $\sim 0.1 (r/100 M)^{-1}$ of the primary signal on average.
\end{abstract}

\maketitle

\section{Introduction}\label{s:intro}
Gravitational wave (GW) observations are expected to open a new
window on the universe within the decade.
First generation GW detectors
(LIGO, Virgo, TAMA, GEO) have been successfully commissioned, and the development of the next
advanced-sensitivity ground based detectors (Advanced LIGO, Advanced Virgo, KAGRA) is well underway.
However, the construction of the planned space-based detector LISA/NGO may be delayed until after 2020.\footnote{{\url{http://lisa.nasa.gov}}, ~~~ {\url{http://www.elisa-ngo.org}}}

It is well known that the GW frequency\footnote{We adopt geometrical units $\G={\rm c}=1$, and
suppress the $\G/\c^2$ and $\G/\c^{3}$ factors to convert mass to length and time units.}, generated by a purely gravitational process
involving mass $M$, typically does not exceed $M^{-1}$.
In particular, for quasicircular black hole (BH) inspirals, the spectrum
peaks near twice the orbital frequency. As the separation shrinks, the GW frequency increases gradually to a
maximum of order $f_{\rm isco} = (6^{3/2} \pi M)^{-1} = 4.4\, M_6^{-1}\,$mHz (where $M_6=M/(10^6 \Msun)$)
corresponding to the innermost stable circular orbit. Then the BHs fall in quickly and the GW frequency reaches
the fundamental quasinormal ringdown frequency $f_{\rm qnr} = (5.4 \pi M)^{-1} = 12\, M_6^{-1}\,$mHz.
Similarly, hyperbolic, parabolic, eccentric, and zoom--whirl encounters have a spectrum that peaks near
$f_{\rm p}\sim (2\pi r_{\rm p}/v_{\rm p})^{-1}<10\,M_6^{-1}\,$mHz, where $r_p\geq 3\, M$ and $v_p<1$
are the distance and relative speed at closest approach. These numbers correspond to
nonspinning BHs, and vary by a factor of $\sim 10$ for spinning binaries \citep{2009CQGra..26p3001B}.
This suggests that GWs are in the sensitive frequency band of terrestrial instruments (10--1000\,Hz)
only for low mass objects below $M_{\max}\sim 10^{4}\,\Msun$ \citep{1998PhRvD..57.4535F,2010ApJ...722.1197A}.

While it is clear that LISA would be very sensitive to GWs emitted by SMBHs, we raise here the possibility whether
ground based GW instruments could also detect massive sources.
Since the instantaneous GW power output during a merger of compact objects
is a constant multiple of $c^5/G\sim10^{52}{\rm W}$,
independent of $M$, which persists for a long time of order $\Delta t \gsim 10\, M = 50\, M_6 \,{\rm s}$,
it is not obvious {\it a priori}, whether a small portion of this enormous GW power { leaking to high frequencies
may be} nonnegligible compared to the detector sensitivity level.

{ We emphasize that the common assumption that black holes
spacetimes cannot radiate at high frequencies $f\gg M^{-1}$ is not true in general. In this paper,
we review previous results that show SMBHs are,
at least in principle, capable of emitting or perturbing high frequency GW signals that LIGO/VIRGO may detect.
We present general arguments that can be
used to search for high frequency SMBH waves in the actual LIGO/VIRGO data, and motivate further
theoretical research to identify astrophysical mechanisms driving such modes. }

High frequency spectral tails, at frequencies well beyond the characteristic scale, $M^{-1}$, may be expected to follow certain
universal trends. A purely gravitational process involving BHs has only one
dimensional scale: the mass, $M$. The mass scaling of the GW spectrum follows from
dimensional analysis. The Fourier transform of the dimensionless GW strain for a source at distance $D$
is most generally
\begin{equation}\label{e:dimensional}
\tilde{h}(f)=\frac{M^2}{D} \psi(M f),
\end{equation}
where $\psi$ is an arbitrary dimensionless
function which may depend on dimensionless source parameters.\footnote{To ensure a finite GW energy requires $\psi( M f ) < (M f)^{-1/2}$ as $f\rightarrow\infty$.}
We are interested in whether there is any
circumstance where the spectrum is an increasing function of $M$ at fixed frequency $f\gg M^{-1}$
if fixing all dimensionless parameters. Based on Eq.~(\ref{e:dimensional}), this is
possible if $\tilde{h}(f)\propto f^{-\alpha}$ at high frequencies for $\alpha\leq 2$.

Similar effects, producing extended powerlaw spectral tails, are common in other fields of science and are
often related to discontinuities in the signal or its derivatives. Examples are the ``leakage'' in data analysis
when signals are abruptly terminated at the end of the sampling interval \citep{1992nrfa.book.....P}, or
``termination waves'' and ``wind contamination'' in geophysical gravity waves \citep{2000JAtS...57.1473P}. A similar
$1/f^{\alpha}$ spectrum is also produced by stochastic processes in physics,
technology, biology, economics, psychology, language  and even music
(for reviews, see \citealt{1978ComAp...7..103P,2002physics...4033M}).
Here, we demonstrate that spectral GW tails associated with (approximate) discontinuities
have a universal high frequency tail, $f^{-\alpha}$, where $\alpha$ is a positive integer, independent of
the details of the waveform. Among these spectral profiles, Eq.~(\ref{e:dimensional}) shows that
an increasing function of mass requires $\alpha= 1$ or 2.

{
The distinct universal shape of these high frequency spectral tails allows to conduct searches for these
GW burst sources irrespective of their origin. The relative arrival times and amplitudes of such signals
in a network of detectors may be used to reduce false alarms and locate the source on the sky.}

If SMBHs were detectable, this would greatly increase the scientific reach of terrestrial GW
instruments. Embedded in gas, some SMBHs shine as bright active galactic nuclei (AGN),
detectable with electromagnetic observatories.
GW detections from SMBHs could serve to trigger searches for EM counterparts { \citep{2005ApJ...629...15H,2006ApJ...637...27K}}, { similar to counterpart searches for merging neutron stars \citep{2010ApJ...725..496N,2012arXiv1210.6362N,2012A&A...539A.124L,2012A&A...541A.155A}}.
An EM counterpart could be used to constrain cosmology { \citep{1986Natur.323..310S,2006PhRvD..74f3006D,2008PhRvD..77d3512M}, extra dimensions \citep{2007ApJ...668L.143D}} and the mass of the
graviton \citep{2008ApJ...684..870K}. High frequency features, sensitive to small scale variability of the spacetime around SMBHs,
may constrain the model of gravity.

In this paper, we provide arguments which may guide future investigations to identify
{ astrophysical} mechanisms involving SMBHs
that may produce high frequency GW spectra for LIGO/Virgo detections. First, we highlight the general connection
between spectral tails and discontinuities using Watson's lemma. Then, we discuss three examples
{ where high frequency leakage may occur}: ultrarelativistic encounters,
quasinormal ringdown modes, and GW echos. { We highlight the need for further research
in this direction to make more detailed assesments of these and other possibilities generating high
frequency spectral tails.}

\section{Theorem on spectral tails}\label{s:watson}
Watson's lemma states that any
piecewise infinitely differentiable, asymptotically exponentially decaying function, $h(t)$,
with a discontinuity in the $n^{\rm th}$ time-derivative at an instant $t_0$,
has an asymptotic spectral tail\footnote{ To avoid confusion, note that the high frequency GW tail discussed here is a \emph{spectral} feature which has nothing to do with the late-time temporal GW tail of quasi-normal modes due to the scattering of GWs off the curved background \citep{1972PhRvD...5.2419P,2008CQGra..25g2001G}.}
$|\tilde{h}(f)|\propto f^{-(n+1)}$ for $f\rightarrow \infty$ \citep{watson}.

Watson's lemma can be generalized to smooth waveforms, which do not have an exact discontinuity in
any derivative, but for which the $n^{\rm th}$ derivative undergoes a rapid transition within (comoving) time
$\Delta t_0\ll M$.
Combining with the dimensional analysis argument, Eq.~(\ref{e:dimensional}),
\begin{equation}\label{e:Watson}
 |\tilde{h}(f)|\propto \frac{M_z^{1-n}}{\dL f^{1+n}},  \qquad {\rm for~~}(\Delta t_{0 z})^{-1} \gg f\gg M_z^{-1},
\end{equation}
where for a source at cosmological redshift $z$, luminosity distance $\dL$,
$M_z=(1+z)M$, $\Delta t_{0 z}=(1+z)\Delta t_0$, and $f=f_{\rm em}/(1+z)$ are the observed (cosmological-redshifted)
quantities where $f_{\rm em}$ is the local emission frequency near the source. The dimensionless constant of
proportionality is independent of $M$, $\dL$, and $f$.

We conclude that the spectral amplitude may be an increasing function of the source mass $M$ at high frequencies
for fixed dimensionless parameters if $n\leq 1$,
i.e., when the signal amplitude or its derivative undergoes a rapid transition within a time $\Delta t_{0z}\ll 1/f$. In the following
we present specific examples where this may be expected.

\section{Gravitational Bremsstrahlung}\label{s:bremsstrahlung}
Gravitational bremsstrahlung during ultrarelativistic flybys of compact objects with a high Lorentz factor $\Gamma$
is an example where the time-domain GW signal exhibits an approximate discontinuity in the first derivative \citep{1978ApJ...224...62K}.
The previous dimensional arguments suggest that ground based instruments may be sensitive to GWs from such
high speed flybys of objects near SMBHs. How large does the relative velocity have to be for detection?

\begin{figure}
\centering{
\mbox{\includegraphics[width=8.5cm]{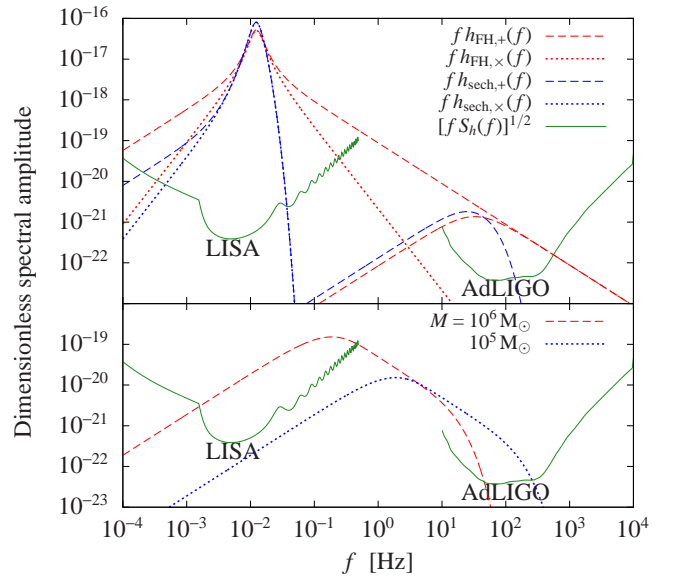}}\\
}
\caption{\label{f:ringultra}
Characteristic spectral amplitude for two waveform families.
{\it Top:} Quasinormal ringdown at distance $100\,$Mpc for mass $M=10^6\Msun$,
amplitude $\A=0.1\tau$, and decay time $\tau/M=10$ (left) and $0.001$ (right curves), respectively.
Red and blue curves correspond to different conventions at $t\leq 0$ (see \S~\ref{s:ring}). { Red and blue curves
represent driving processes that terminate abruptly and gradually, respectively.}
{\it Bottom:} Ultrarelativistic flybys with
 $\Gamma=50$, impact parameter $b=20\,M$, and mass ratio $q = 10^{-5}$ for $M=10^6$ (red) and $10^5\,\Msun$ (blue) at $1\Gpc$.
Solid green curves show the sensitivity of LISA (left) and Advanced LIGO (right).
{ Advanced LIGO may detect} weakly damped SMBH ringdown modes excited by a process that terminates abruptly
and ultrarelativistic encounters.
}
\end{figure}

The GWs have been calculated for ultrarelativistic encounters where  $b$, the impact parameter, is sufficiently large
to ensure that the two objects move on nearly linear trajectories \citep{1978ApJ...224...62K}.
The time domain GW signal
is a ``v-shaped'' burst with a sharp discontinuity in the first derivative and
is beamed in the direction of motion.
In the restframe of mass $m_1$,
\begin{eqnarray}
 \tilde{h}(f) &\approx& 4 \frac{ m_1 m_2}{\dL}\frac{(1+v^2) \Gamma}{v^2}\left[(1-v)K_2(w) - (1+v)K_2(u)\right]\nonumber\\
&\approx&
\frac{8}{\pi^2} \frac{q}{ \bar{b}^2 \dL  } \frac{\Gamma^{3}}{f^2},{\,\rm if\,}
\frac{\Gamma}{2\pi b}\lesssim f \lesssim \frac{\Gamma^2}{2\pi b} {\rm~and~} \Gamma\gg 1,
\label{e:kt1}
\end{eqnarray}
in the forward direction at angles $\theta\ll \Gamma^{-1/2}$, where $\Gamma=(1-v^2)^{-1/2}$ is the Lorentz factor,
$v$ is the relative velocity.
Here, $K_2(x)$ is a modified Bessel function, $u=2\pi f b/(v\Gamma)$, $w=2\pi f b/[v(1+v)\Gamma^2]$, $q=m_2/m_1<1$
is the mass ratio, $b$ is the impact parameter, and $\bar{b}=b/m_1$.
Assuming that $m_1\approx M \gg m_2$, self-consistency requires $\bar{b}\gg 1$
{ (see \citealt{1978PhRvD..18..990D,1992PhRvD..46..694D,2001PhRvD..64f4018S},
and discussion below for the limit $b\sim 0$)}. We find that the signal extends into the
Advanced LIGO/Virgo frequency band ($f\gtrsim 10\,$Hz) if $\Gamma \gtrsim 18\, M_6^{1/2} \bar{b}^{1/2}$.
Eq.~(\ref{e:kt1}) shows that the spectral amplitude is independent of $M$ for fixed dimensionless
parameters and fixed $f$, as it should, based on the dimensional arguments of Eq.~(\ref{e:Watson}).
Figure~\ref{f:ringultra} (bottom panel) shows that the amplitude may well exceed the Advanced LIGO/Virgo
sensitivity level for SMBHs at 1\,Gpc.

{ Whether or not such ultrarelativistic objects exist near supermassive black holes
is unknown at present, but we argue that this possibility cannot be ruled out on theoretical grounds and
may be tested observationally with Advanced LIGO and VIRGO.}
Ultrarelativistic emission with Lorentz factors between $\Gamma \sim 5$ to $40$
is observed in active galactic nuclei with jets
\citep{1995PASP..107..803U,2001ApJS..134..181J,2005ApJ...625...72S,2005AJ....130.1418J}.
{ These jets are believed to be launched by electromagnetic processes near a Kerr black hole
\citep{1977MNRAS.179..433B,2009MNRAS.394L.126M}. The total mass outflow rate during flares, inferred from observations,
reaches values comparable to the mass accretion rate of the BH, up to $\Msun/{\rm yr}$ \citep{2008MNRAS.385..283C}.
These observations suggest that at least energetically,
the magnetic field and angular momentum of a Kerr SMBH is sufficiently large
to accelerate a neutron star or white dwarf to ultrarelativistic velocities.
\citet{2001PhRvD..64f4018S} considered the high frequency GW signal during
the acceleration of a stellar mass localized blob of matter.
In this case, $h(t)\sim 4 \Gamma^3 m  / \dL$ in the
forward direction and $2 \Gamma m (1+\cos \theta) / \dL$ at large angles $\theta \gg \Gamma^{-1}$. Here
$m$ is the mass of the accelerated object, and the frequency extends to $f_{\rm cut}\sim\Gamma^2/\Delta t$,
where $\Delta t$ is the acceleration timescale in the SMBH restframe. }

{
Secondly, a merging SMBH binary may gravitationally accelerate stellar mass compact objects
to ultrarelativistic velocities, if they happen to be in the vicinity of the merging binary \citep{2010ApJ...711L..89V}.
The latter possibility is supported by several arguments. First, a widely separated
SMBH binary gravitationally traps stellar objects in mean motion resonances from the stellar cluster
surrounding the binary. These objects are analogues of Trojan satellites in the Solar System corotating with Jupiter.
An accretion disk around the binary may also supply objects to mean motion resonances
by capturing stars crossing the disk or
forming stars in the outskirts of the disk and transporting them radially inwards close to the binary
\citep{1999A&A...352..452S,2001A&A...376..686K,2012MNRAS.423.1450A}.
As the SMBH binary inspirals, objects locked in a mean-motion resonance are dragged inwards
to small radii down to $10$ Schwarzschild radii \citep{2010ApJ...724...39S,2010PhRvD..81j3004S,2011MNRAS.415.3824S}.
Numerical relativity simulations by \citet{2010ApJ...711L..89V} have shown that test particles surrounding a SMBH
binary may be accelerated to high Lorentz factors during the final merger of the binary. If so, the
GW bremsstrahlung of these objects due to the SMBH
may be candidates for LIGO/VIRGO detections. Future studies should investigate this process in more detail
for a variety of initial conditions, SMBH binary mass ratios and spins, and stellar mass gravitating particles 
to determine the likelihood of such detections.}

\section{Black Hole Ringdown}\label{s:ring}

A perturbed BH behaves as a damped harmonic oscillator, and emits GWs \citep[for a review, see][]{2009CQGra..26p3001B}
\begin{equation}
\left(
\begin{array}{l}
 h_{+}(t)\\
h_{\times}(t)
\end{array}
\right)
 = \sum_k \frac{\A_k S_{2,k} }{\dL}  e^{-t/\tau_k}
\left(
\begin{array}{l}
\cos(2\pi f_{k} t)\\
\sin(2\pi f_{k} t)
\end{array}
\right)\,, \label{e:ring1}
\end{equation}
where $+$ and $\times$ denote the polarization,
$t>0$ is time from the end of the external perturbation,
$f_k$ is the quasinormal ringdown frequency,
$1/\tau_k$ is the decay rate,
and $S_{2,k}$ are spin-2 weighted spheroidal harmonics
which { set the angular distribution of the emission} with root mean square $(4\pi)^{-1/2}$.
Here $k$ is a discrete index which runs over all of the ringdown modes
$(n,\ell,m)$ analogous to the principal, angular momentum, and magnetic quantum numbers of the hydrogen atom.
While $f_k$ and $\tau_k$ are universal properties of BHs which depend only on their mass and spin, the
$\A_k$ amplitude coefficients, called quasinormal excitation factors, are sensitive to the details of the perturbation. The excitations factors can be computed as a convolution of an external source with the Green's function of the SMBH \citep{1986PhRvD..34..384L,1995PhRvD..51..353A,2006PhRvD..74h4028D,2006PhRvD..74j4020B,2007PhRvD..75l4024B,2010PhRvD..81j4048B,2011PhRvD..84j4002D,2011PhRvD..84d7501H}. 

To compare with the sensitivity levels of GW instruments, one needs to calculate $\tilde{h}(f)$, the Fourier transform of the waveform.
For a damped harmonic oscillator, the high frequency tail depends on
the start of the waveform. Two conventions are used in the literature discussing the detectability of ringdown
waveforms \citep{1989PhRvD..40.3194E,1992PhRvD..46.5236F,1998PhRvD..57.4535F,2007PhRvD..76j4044B,2006PhRvD..73f4030B}.
In the {\it Echeverria-Finn} (EF)
convention, the signal is assumed to vanish for $t< 0$ and follow Eq.~(\ref{e:ring1}) for $t\geq 0$. In the
{\it Flanagan-Hughes} (FH) convention, to avoid an unphysical discontinuity at $t=0$, the signal is extrapolated
symmetrically for $t<0$ and the overall amplitude is renormalized by $2^{-1/2}$ to keep the total
GW energy fixed. However, the signal used in the FH convention is also discontinuous in the first derivative,
which has an impact at high frequencies.
{ EF and FH introduced these conventions for convenience to calculate
the detectability near the characteristic ringdown frequency of the signal, but here we explore the high frequency asymptotics of these conventions
and identify the corresponding physical conditions. }
Physically, an EF spectrum is appropriate in the case where the QNM is excited by an instantaneous pulse.
The FH spectrum represents excitation processes that build up the mode continuously, but generate 
a sudden change in the derivative of the waveform as the excitation is terminated abruptly at $t=0$, 
letting the BH ring freely thereafter.
For waveforms excited smoothly on timescales $\tau_k$, we introduce a third convention which
is continuous in all derivatives, by replacing $e^{-t/\tau_k}$ by $\sqrt{2}\, \sech(t/\tau_k)=\sqrt{8}/(e^{t/\tau_k}+e^{-t/\tau_k})$
in Eq.~(\ref{e:ring1}).

The Fourier transform of the ringdown waveform can be obtained analytically
in the three conventions for each mode $k$ { comprising the waveform, independently}. In the limit of
asymptotically large frequencies $f\gg \max(f_k,\tau_k^{-1})$, we get
\begin{eqnarray}
 &&|\tilde{h}_{k+}^{\EF}| \approx  \frac{\A_k S_{2,k} }{2\pi \dL f},\quad
|\tilde{h}_{k\times}^{\EF}| \approx  \frac{\A_k S_{2,k} f_k}{2\pi  \dL f^2},\label{e:asympt1}\\
 &&|\tilde{h}_{k+}^{\FH}(f)| \approx  \frac{\A_k S_{2,k} }{\sqrt{8}\pi^2 \tau_k \dL f^2},\quad
|\tilde{h}_{k\times}^{\FH}(f)| \approx  \frac{\A_k S_{2,k} f_{k}}{\sqrt{8}\pi^2 \tau_k \dL f^{3}},\\
 &&
\left(
\begin{array}{l}
\tilde{h}_{k+}^{\sech}(f)\\
\tilde{h}_{k\times}^{\sech}(f)\\
\end{array}
\right)
 \approx  \frac{\pi}{\sqrt{2}}\frac{\A_k S_{2,k} \tau_k}{\dL \cosh(\pi^2\tau_k f)}
\left(
\begin{array}{l}
\cosh(\pi Q_k)\\
\sinh(\pi Q_k)\\
\end{array}
\right)
\label{e:asympt4}
\end{eqnarray}
where $Q_k\equiv \pi \tau_k f_k$ is the quality factor of the damped oscillator.
Equations~(\ref{e:asympt1}--\ref{e:asympt4}) show that the asymptotic spectral tail
may enter the detector band even if $f\gg M^{-1}$.
For fixed $f$, the spectral GW amplitude  scales very differently with $M$.
Since $\A_k$, $\tau_k$ and $f_k^{-1}$ are all proportional to $M$ for fixed $k$,
we find that $|\tilde{h}^{\EF}_{k +}(f)|\propto M$,  $|\tilde{h}^{\EF}_{k \times}(f)|$ and $|\tilde{h}^{\FH}_{k +}(f)|$ are
independent of $M$, and $|h^{\FH}_{k \times}|\propto M^{-1}$, while
$|\tilde{h}^{\sech}_{k \times}(f)|$ and $|h^{\sech}_{k +}(f)|$ are asymptotically exponentially suppressed for $f\gg M^{-1}$.
In the later case,  the peak spectral amplitude is proportional to $\A_k \tau_k\propto M^2$ for a fixed $k$
at frequencies $f\sim \tau_{k}^{-1}$.
At $f\sim \tau_k^{-1}$ for some $k$, the orientation-averaged one-sided dimensionless GW spectral amplitude per logarithmic
frequency bin for this signal is $\langle 2f \tilde{h}^{\sech}_{+}\rangle \sim (4/5)^{1/2} (4\pi)^{-1/2} (2^{-1/2}\pi) \, \A_k/\dL$.
These mass scalings also follow from dimensional analysis once the spectral shape is specified (\S~\ref{s:intro})
and are consistent with the general expectations from Watson's lemma
(\S~\ref{s:watson}).
The EF and FH spectra become invalid beyond a frequency $f\gtrsim \Delta t^{-1}$, where $\Delta t$ represents
the characteristic timescale over which the excitation occurs generating the approximate discontinuity
in the signal amplitude or derivative, respectively.

The top panel of Figure~\ref{f:ringultra} { shows the angular averaged spectral amplitude for $M=10^6\,\Msun$, $d_L=100\,$Mpc, and $\A_{k}=0.1\tau$. Different line types correspond to different conventions at the start of the waveform as labelled. Two
sets of curves peaked at $\sim 0.01$ and $30\,$Hz represent $\tau=10\,M$ (weakly damped modes) and $10^{-3}M$ (highly damped modes), respectively.}
The GW signal leaks into the LIGO/Virgo { frequency} band for FH-type excitation
processes both for weakly and highly damped modes.
If the excitation occurs smoothly over timescale $\tau$ (blue curves), then the signal may enter the LIGO/Virgo
band if $\tau=10^3 \Msun$, but not if $\tau$ is much larger.\footnote{Note that the $\times$ polarization of the waveform is suppressed at high frequencies because the signal vanishes at $t=0$ in this case, see Eq.~(\ref{e:ring1}). Only the $+$ polarization is detectable at high frequencies for smooth excitation processes for highly damped modes (blue dashed curve).}
Comparing to the minimum frequency and sensitivity level of Advanced LIGO/VIRGO,
{\it we conclude that SMBH ringdown waveforms may be detectable on average\footnote{averaging over polarization and direction} to $\sim 100\,{\rm Mpc}$ if the quasinormal modes are excited to an amplitude exceeding
$\A_{k}\gtrsim 0.01\,\tau$.}

Going beyond the statement that some QNMs { may} have a potentially detectable high frequency tail,
it would be desirable to identify conceivable { astrophysical processes} exciting these modes to detectable levels.
The $\A_{k}$ excitation factors of QNMs are explored only for a very limited number of configurations to date
\citep{1986PhRvD..34..384L,1995PhRvD..51..353A,2006PhRvD..74h4028D,2006PhRvD..74j4020B,2007PhRvD..75l4024B,2010PhRvD..81j4048B,2011PhRvD..84j4002D,2011PhRvD..84d7501H}.
{ For this discussion, we distinguish three classes of ringdown modes.
\begin{itemize}
 \item  High--$\ell$ modes. For of a Schwarzschild BH,
$f_{n \ell m}\approx  \ell \Omega_c/(2\pi)$ and $1/\tau_{n \ell m} \approx n \Omega_c$, where
$\Omega_c=27^{-1/2} M^{-1}$ is the orbital frequency at the light ring \citep{1971ApJ...170L.105P}.\footnote{
Similar expressions hold for Kerr BHs, see \citet{1984PhRvD..30..295F,2012arXiv1207.4253Y,2012arXiv1207.2460K}.}
These modes may be identified with energy leakage from the photon sphere  \citep{1972ApJ...172L..95G,1984PhRvD..30..295F,2011PhRvD..84j4002D,2012arXiv1207.4253Y}.
They ring in the LIGO/VIRGO band if $10\,{\rm Hz}\lesssim f_{n \ell m} \lesssim 1000\,{\rm Hz}$.
 \item Weakly damped modes. Consider modes with $f_{n \ell m}\lesssim M^{-1}$, $1/\tau_{n \ell m}\sim M^{-1}$.
Although, the ringdown frequency is well outside the LIGO/VIRGO frequency range
in this case for large $M$,
a rapidly changing external driving process on timescale $\Delta t \lesssim 0.1\,{\rm s}$ may lead to high frequency tails for these modes (see e.g. EF and FH above). This is analogous to driving a harmonic oscillator with a process that has a component well above the resonant frequency. Further studies are necessary to see if the high frequency spectral amplitude of the QNM may be sufficiently large for any specific driving process in practice.
 \item Highly damped modes. The decay rate of very high overtone modes may be arbitrarily large,
$\lim_n 1/\tau_{n \ell m}= \infty$ for all $\ell$ and $m$, which implies a high frequency spectral tail.
Although the quality factor of these modes approaches zero, their excitation may still be sufficient to reach
levels required for detection as stated above ($\A_k\gtrsim 0.01\tau$ at 100\,Mpc).
For a damped oscillator driven by a stationary external process of amplitude less than unity,
$\A_k\lesssim \tau_k / (M f_k)$. These modes are detectable to 100\,Mpc with Advanced LIGO if
$f_k\lesssim 100\,M^{-1}$ and $\tau\sim 0.1\,{\rm s}$ (see the right blue curve on the top panel of Figure~\ref{f:ringultra} and discussion above),
which is satisfied for all modes with $\ell\lesssim 3000$.
\end{itemize}
}
{ We highlight two astrophysical processes which may generate high frequency QNMs.
First, a merging SMBH binary gravitationally accelerates particles
and nearby stellar mass compact objects to ultrarelativistic velocities which may then escape
or merge with the SMBH (\citealt{2010ApJ...711L..89V}, and see \S~\ref{s:bremsstrahlung} above).}
Ultrarelativistic collisions { with a SMBH} \citep{1972PhRvL..28.1352D,2010PhRvD..81j4048B}
excite high-$\ell$ ringdown modes.
The radiated energy scales as $E_{\ell}\propto 1/\ell^{-d_E}$, where $d_E\sim 2$ for radial infall
and $1$ for grazing captures, implying that $\tilde{h}(f)\sim f^{-3/2}$ to $f^{-2}$ for high $f$.
{ Thus, similar to ultrarelativistic gravitional bremsstrahlung,
the high frequency spectral GW amplitude of ultrarelativistic capture sources is a constant or increasing function of
mass for a fixed mass ratio (see Eq.~\ref{e:dimensional}). }

{ Secondly, the high frequency gravitational effects of a stellar (or intermediate) mass ($M_b$) black hole binary
may excite high frequency QNMs of the SMBH. Stellar mass compact binaries form during
close encounters and merge due to GW emission \citep{2009MNRAS.395.2127O,2012PhRvD..85l3005K},
or they may be driven to merger due to the Kozai effect
\citep{2003ApJ...598..419W,2012ApJ...757...27A,2012arXiv1206.4316N}.
The QNM excitation function by these sources has a high frequency component and the driving force
terminates on a short timescale proportional to $M_b$. Based on the arguments outlined above,
it is then plausible to expect a high frequency ringing of the SMBH.
Furthermore, the high frequency GWs emitted by such  stellar  (or intermediate) mass compact binaries
are scattered by the SMBH, generating GW echos. The scattered waves represent high-$\ell$ ringdown modes, which we discuss in \S~\ref{s:scattering} below.
Further studies are necessary to examine the excitation factors and the
detectability of the signal  to assess the likelihood of detecting these waveforms}.

\section{GW lensing echos and diffraction}\label{s:scattering}
{ Consider the GWs emitted by an inspiraling stellar mass BH binary
source in a galactic nucleus near a SMBH.}
High frequency GWs are scattered by the gravitational field of the SMBH \citep[e.g.][]{1988sfbh.book.....F}
producing GW echos \citep{2011PhRvD..84j4002D,2012PhRvD..86f4030Z}.
The GW frequency is not effected to leading order in the WKB limit if
the frequency of the incoming wave exceeds $M^{-1}$.
The scattered GWs, or GW echos, are completely analogous to the relativistic images of electromagnetic waves
for rays going around the SMBH in opposite directions and/or multiple times.
{ However, GW echos may be easier to distinguish from astrophysical foregrounds than
the relativistic images of EM sources since high frequency GW sources
are sufficiently rare, have a short duration, and a distinct frequency evolution \citep{2009MNRAS.395.2127O,2012PhRvD..85l3005K}.}
Furthermore, GW detectors are sensitive to the
signal phase and amplitude, i.e. square-root of the intensity, { which is suppressed much less than
the EM intensity} (see Fig.~\ref{f:lensing} below).

The GW echo arrives at the detector after the primary signal with a delay corresponding to the projected light-travel
time between the object and the source,
\begin{equation}\label{e:DeltaT}
\Delta T=( 1 - \cos \alpha) r \sim 14{\,\rm h}\times ( 1 - \cos \alpha) M_6 (r/10^4 M),
\end{equation}
where $r$ is the distance between the binary and the SMBH and $\alpha$ is the deflection angle.
Note, that the probability distribution of inspiraling stellar mass BH binaries  in galactic nuclei
formed by GW emission during close encounters
increases towards the SMBH { \citep{2009MNRAS.395.2127O}}.
A signficant fraction ($\sim 30\%$) of sources are within $r\lesssim 10^4 M$ for which the GW echos occur
within hours of the primary signal.

We calculate the amplitude of the GW echo for a source at radius $r$ from the SMBH
in the geometrical optics limit using the \citet{2000PhRvD..62h4003V}
lens equation valid for arbitrary deflection angles.
{ We direct the reader to that paper for details \citep[see also][]{2008PhRvD..78j3005B}.}
For a sanity check, we also calculate the scattering cross-section of plane waves
impinging on the SMBH, which approaches the lensing cross-section in the large
deflection angle limit if $r\gg M$.
The amplitude of the GW echo may be larger or smaller than the primary GW signal of the source depending
on the relative position of the source, the SMBH, and the line-of-sight as shown by Figure~\ref{f:lensing}.
The lensing effect may strongly increase the intensity of the wave if the lens
and the source are nearly colinear along the line of sight \citep{1936Sci....84..506E}.
For $r\ll D$, the deflection angle corresponding to the Einstein radius
for strong magnification is $\alpha_{\rm E}\sim (4r/M)^{-1/2}$, i.e. $(11^{\circ},3.6^{\circ}, 1.1^{\circ})$
for $r=(10^2,10^3,10^4)M$, respectively. For an isotropic distribution of sources, this implies that
$(2,0.2,0.02)\%$ of sources are within the Einstein radius for the three values of $r$, respectively.
For larger deflection angles, the GW amplitude decreases as $h\propto h_0/(r\alpha^{2})$, where
$h_0$ is the GW amplitude without lensing.
The peak at $\alpha \sim \pi$ (parallel backscattering)
is known as the glory effect of geometrical optics \citep{1988sfbh.book.....F}.
The figure shows that the amplitude of relativistic images may be significant
even for large deflection angles if $r\lesssim 10^3 M$ and the primary signal
has an SNR larger than $\sim 50$.

\begin{figure}
\centering{
\mbox{\includegraphics[width=8.5cm]{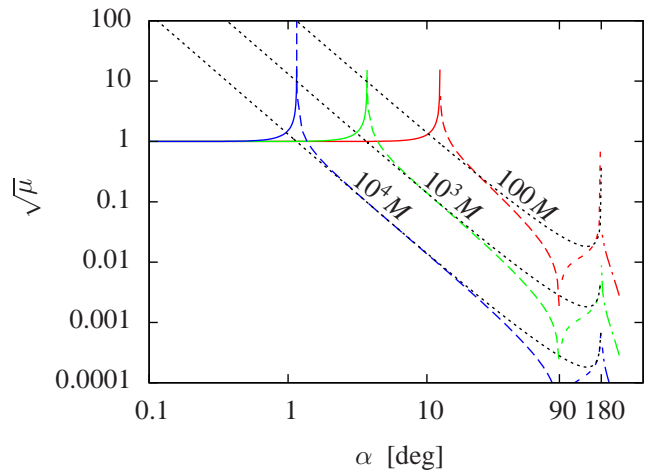}}\\
}
\caption{\label{f:lensing}
Gravitational lensing magnification of a high frequency GW source by a SMBH as a function of deflection angle $\alpha$.
Colored curves use the thin lens equation, black dotted lines show the plane wave limit.
The three sets of curves correspond to different source distances $r$ from the SMBH as labeled,
different line styles represent relativistic images of different order.
Solid curves show rhe primary signal; dashed and dotted-dashed curves 
show the amplitude of the first GW echo.
The peak at small deflection angles corresponds to strong lensing at the Einstein radius $\alpha_E = (4r/M)^{-1/2}$,
the second peak at $180^\circ$ represents the glory phenomenon
in geometrical optics. Additional peaks appearing at $\alpha=n \pi$, where $n$ is an integer, are not shown beyond $n=1$.
The GW frequency is assumed to greatly exceed $M^{-1}$.
Diffraction fringe patterns (not shown) appear with spacing
$\delta \alpha \sim (34\,M f)^{-1}$ and sharper features are smoothed.
}
\end{figure}

Figure~\ref{f:lensing} assumes $f\gg M^{-1}$ and neglects diffraction effects
due to the interference of rays going around the SMBH in opposite directions and
the QNM of the SMBH \citep{1985PhRvD..31.1869M,1998PhRvL..80.1138N}. This is a good approximation
for typical large deflection angles of stellar mass GW inspirals at the typical distances $r\gg 100 M$,
since the timescale for the increase of the GW frequency is
much less than $\Delta T$ (Eq.~\ref{e:DeltaT}). However, interference effects may become significant for
sources that happen to be behind or in front of the SMBH close to the  line of sight. This
smears out the magnification peak and produces a GW fringe pattern in amplitude and phase
on the angular scale $\delta\alpha \sim 1/(34 f M) \sim 5.9\times 10^{-4} \times M_6^{-1} (f/10\Hz)^{-1}$ { for
a nonspinning SMBH (similar expressions with different prefactors hold for spinning SMBHs, see \citealt{2008CQGra..25w5002D})}.

{ The frequency and direction dependence of the
GW diffraction patterns introduce interesting time variations in the signal.
The diffraction pattern scale shrinks in time for inspiraling binary sources
as $f$ increases \citep{2008CQGra..25w5002D}.
Furthermore,} the relative motion of the lens and the source lead to a varying GW amplitude
analogous to the microlensing events of electromagnetic signals \citep{1991ApJ...374L..37M}
and phase shift.
Assuming that the source moves along a circular orbit around the SMBH with
a Keplerian angular velocity, $\Omega_{\rm K}$, it travels across the diffraction fringe within
$\delta t_f =\delta \alpha /\Omega_{\rm K} \sim 9.3\,{\rm s}\times(r/10^3 M)^{3/2} (f/100\,{\rm Hz})^{-1}$.
This is much less than the source lifetime for inspiraling stellar mass binaries.
Based on the strong magnification rates { estimated above}, we conclude that a small fraction of sources ($\lesssim 2$\%)
might exhibit these features.

{ If measured, the GW echo may be used to learn about the parameters of the SMBH beyond its
distance from the merging stellar-mass binary.
Time dependent effects of GW diffraction patterns of an inspiral signal and/or the
microlensing effect discussed above carry information about the mass of and spin of the SMBH. The polarization of the
scattered waves relative to that of the primary signal is also sensitive to the SMBH spin \citep{2008CQGra..25w5002D}. }

\section{Discussion}
Ground based instruments may, in principle, be sensitive to GWs
emitted by SMBHs with arbitrarily large $M$ at frequencies $f\gg M^{-1}$. Dimensional analysis
shows that the spectrum $\tilde{h}(f)$ is a growing function of mass or  mass-independent at large
frequencies if $\tilde{h}(f)\propto f^{-1}$ or $f^{-2}$, respectively.
Watson's lemma shows that such a spectrum corresponds to a waveform having an abrupt change in amplitude
or in its first time-derivative within a short time ($\Delta t< 1/f$).
Discontinuous GW shock fronts with arbitrarily small $\Delta t$
are self-consistent, exact solutions of general relativity (\citealt{1951PhRv...83...10V,1971GReGr...2..303A};
and see also cosmic strings for which $\tilde{h}(f)\propto f^{-1/3}$, \citealt{2000PhRvL..85.3761D,2005PhRvD..71f3510D}),
so one should not discard this possibility a priori.

{Ground-based GW detectors could provide observational constraints on the event rates of such
sources independent of their origin. Unlike other GW burst sources, these   sources have a well-defined broadband spectral
shape, $\tilde{h}(f)\propto f^{-1}$ or $f^{-2}$, allowing to develop optimal search algorithms for their
efficient detection. These signals may be distinguished from instrumental glitches if it shows up coincidentally
in all of the different instruments in a detector network (i.e. the two LIGO sites, Virgo, and Kagra) 
with a consistent spectral amplitude.}

We have discussed three { hypothetical examples involving SMBHs potentially capable of generating} high-frequency GW signals:
ultrarelativistic sources, QNM ringdown waveforms excited externally by { high frequency} processes
{ or those} that terminate abruptly, and the scattering and diffraction of GWs.
{ First, an ultrarelativistic stellar mass object moving in the gravitational field of
a SMBH emits gravitational bremsstrahlung.}
The signal derivative undergoes a rapid transition on an observed timescale
$f_{\rm cut}^{-1}\sim\Delta t/\Gamma^2$, where $\Gamma$ is the Lorentz factor and
$\Delta t$ is the acceleration timescale proportional to the impact parameter, $b$.
Then for $M^{-1} \ll f \lesssim f_{\rm cut}$, the GW spectrum $\tilde{h}(f)\propto f^{-2}$ is
{ proportional to the mass of the object but independent of the mass of the SMBH, $M$,
extending well into the sensitive range of ground based GW instruments if $\Gamma \gtrsim 5.7\, M_5 \bar{b}$.
Here $M_5=M/10^5\Msun$ and $\bar{b}=b/M$.}
{ Second}, SMBHs, if perturbed, ring as damped harmonic oscillators, and emit ringdown GWs.
We have shown that { weakly damped ringdown waveforms leak into the LIGO frequency band
if the excitation process has a { high frequency component or} terminates within $0.1\,{\rm s}$.
These GWs may be detectable for}
arbitrary $M\gg 100\,\Msun$, with Advanced LIGO/Virgo-type instruments to $100\Mpc$
{ if the GW amplitude satisfies} $\A_k\gtrsim 0.01\tau$,
{ where $\tau$ is the decay time of the QNM.}
{ However, detailed calcalations of excitation factors are necessary
to determine whether or not this level may be reached for specific realistic configurations.}

Finally, lensing and diffraction effects of high frequency GWs { by SMBHs} may also generate
detectable features. In particular, we expect a $0.02$\%--$2\%$ of LIGO/Virgo sources in galactic nuclei
{ (e.g. inspiral and merger of stellar mass black hole or neutron star binaries)}
could be strongly magnified { by the central SMBH}.
{ Diffraction limits the maximum magnification of these sources.}
{ The rest of the high frequency sources in the vicinity of SMBHs produce weak GW echos due to}
large deflection angle scattering by the SMBH. { The GW echo amplitude is typically between $0.1$--$10\%$
of the primary signal for these sources.} The time delay between the primary and scattered signals in this case
is typically tens of minutes.

It remains to be seen whether these processes occur in reality.
To date, numerical simulations of binary mergers did not find evidence for waveforms having such sharp
features for quasicircular mergers \citep{2007PhRvD..75l4024B}, except for spurious GWs related to the
improper choice of initial conditions \citep{2010PhRvD..81l4030F}. However these simulations are limited
{ by resolution} to
comparable mass binaries (see \citealt{2011PhRvL.106d1101L} for the most unequal mass ratio) and cannot resolve
high order ringdown modes. Recently, \citet{2010PhRvD..81j4048B} carried out simulations for
head-on, grazing, and off-axis collisions with nonzero initial velocity. They found that the cutoff frequency is shifted to
larger values for higher Lorentz factors and multipole moments.
Future studies should investigate other possible examples, involving SMBHs.
These may include high frequecy ringdown resonances in extreme mass ratio inspirals or
three-body encounters.
High frequency GWs of SMBHs, may also arise in alternative theories of
gravity or cosmology due to the self-gravity of GWs
\citep{2003PhRvD..68d4022F,2005PhRvD..71h6003K}.

One should also be aware of these spectral effects when constructing time domain templates
matching the inspiral, merger, and ringdown waveforms \citep{2007PhRvD..75b4005B,2010PhRvD..82f4016S}.
A high-frequency powerlaw spectral tail is sensitive to a discontinuity in the
derivative of the time-domain waveform at the matching boundaries.
We have shown that the high frequency QNMs are very sensitive to the initial conditions for
both the EF and the FH conventions. This theorem may also be useful for debugging numerical codes.

\acknowledgements
I thank Emanuele Berti, Ryan O'Leary, Dan Holz, Lorenzo Sironi, Zoltan Haiman,
and Scott Tremaine for discussions, and Nico Yunes and Cole Miller for useful comments after carefully reading 
the manuscript.
The author acknowledges support by NASA through Einstein Postdoctoral Fellowship grant number
PF9-00063 awarded by the Chandra X-ray Center (operated by the SAO,
contract NAS8-03060).

\bibliographystyle{apj}
\bibliography{apj-jour,ms}

\begin{thebibliography}{81}
\expandafter\ifx\csname natexlab\endcsname\relax\def\natexlab#1{#1}\fi

\bibitem[{{Abadie} {et~al.}(2012){Abadie}, {Abbott}, {Abbott}, {Abbott},
  {Abernathy}, {Accadia}, {Acernese}, {Adams}, {Adhikari}, {Affeldt}, \&
  et~al.}]{2012A&A...541A.155A}
{Abadie}, J., {et~al.} 2012, \aap, 541, A155

\bibitem[{{Aichelburg} \& {Sexl}(1971)}]{1971GReGr...2..303A}
{Aichelburg}, P.~C., \& {Sexl}, R.~U. 1971, Gen. Rel. Grav., 2, 303

\bibitem[{{Amaro-Seoane} \& {Santamar{\'{\i}}a}(2010)}]{2010ApJ...722.1197A}
{Amaro-Seoane}, P., \& {Santamar{\'{\i}}a}, L. 2010, \apj, 722, 1197

\bibitem[{{Andersson}(1995)}]{1995PhRvD..51..353A}
{Andersson}, N. 1995, \prd, 51, 353

\bibitem[{{Antonini} \& {Perets}(2012)}]{2012ApJ...757...27A}
{Antonini}, F., \& {Perets}, H.~B. 2012, \apj, 757, 27

\bibitem[{{Ayliffe} {et~al.}(2012){Ayliffe}, {Laibe}, {Price}, \&
  {Bate}}]{2012MNRAS.423.1450A}
{Ayliffe}, B.~A., {Laibe}, G., {Price}, D.~J., \& {Bate}, M.~R. 2012, \mnras,
  423, 1450

\bibitem[{{Babak} {et~al.}(2007){Babak}, {Fang}, {Gair}, {Glampedakis}, \&
  {Hughes}}]{2007PhRvD..75b4005B}
{Babak}, S., {Fang}, H., {Gair}, J.~R., {Glampedakis}, K., \& {Hughes}, S.~A.
  2007, \prd, 75, 024005

\bibitem[{{Baker} {et~al.}(2007){Baker}, {McWilliams}, {van Meter},
  {Centrella}, {Choi}, {Kelly}, \& {Koppitz}}]{2007PhRvD..75l4024B}
{Baker}, J.~G., {McWilliams}, S.~T., {van Meter}, J.~R., {Centrella}, J.,
  {Choi}, D.~I., {Kelly}, B.~J., \& {Koppitz}, M. 2007, \prd, 75, 124024

\bibitem[{{Berti} {et~al.}(2007){Berti}, {Cardoso}, {Cardoso}, \&
  {Cavagli{\`a}}}]{2007PhRvD..76j4044B}
{Berti}, E., {Cardoso}, J., {Cardoso}, V., \& {Cavagli{\`a}}, M. 2007, \prd,
  76, 104044

\bibitem[{{Berti} \& {Cardoso}(2006)}]{2006PhRvD..74j4020B}
{Berti}, E., \& {Cardoso}, V. 2006, \prd, 74, 104020

\bibitem[{{Berti} {et~al.}(2010){Berti}, {Cardoso}, {Hinderer}, {Lemos},
  {Pretorius}, {Sperhake}, \& {Yunes}}]{2010PhRvD..81j4048B}
{Berti}, E., {Cardoso}, V., {Hinderer}, T., {Lemos}, M., {Pretorius}, F.,
  {Sperhake}, U., \& {Yunes}, N. 2010, \prd, 81, 104048

\bibitem[{{Berti} {et~al.}(2009){Berti}, {Cardoso}, \&
  {Starinets}}]{2009CQGra..26p3001B}
{Berti}, E., {Cardoso}, V., \& {Starinets}, A.~O. 2009, Classical and Quantum
  Gravity, 26, 163001

\bibitem[{{Berti} {et~al.}(2006){Berti}, {Cardoso}, \&
  {Will}}]{2006PhRvD..73f4030B}
{Berti}, E., {Cardoso}, V., \& {Will}, C.~M. 2006, \prd, 73, 064030

\bibitem[{{Blandford} \& {Znajek}(1977)}]{1977MNRAS.179..433B}
{Blandford}, R.~D., \& {Znajek}, R.~L. 1977, \mnras, 179, 433

\bibitem[{{Bozza}(2008)}]{2008PhRvD..78j3005B}
{Bozza}, V. 2008, \prd, 78, 103005

\bibitem[{{Celotti} \& {Ghisellini}(2008)}]{2008MNRAS.385..283C}
{Celotti}, A., \& {Ghisellini}, G. 2008, \mnras, 385, 283

\bibitem[{{Dalal} {et~al.}(2006){Dalal}, {Holz}, {Hughes}, \&
  {Jain}}]{2006PhRvD..74f3006D}
{Dalal}, N., {Holz}, D.~E., {Hughes}, S.~A., \& {Jain}, B. 2006, \prd, 74,
  063006

\bibitem[{{Damour} \& {Vilenkin}(2000)}]{2000PhRvL..85.3761D}
{Damour}, T., \& {Vilenkin}, A. 2000, Physical Review Letters, 85, 3761

\bibitem[{{Damour} \& {Vilenkin}(2005)}]{2005PhRvD..71f3510D}
---. 2005, \prd, 71, 063510

\bibitem[{{Davis} {et~al.}(1972){Davis}, {Ruffini}, {Tiomno}, \&
  {Zerilli}}]{1972PhRvL..28.1352D}
{Davis}, M., {Ruffini}, R., {Tiomno}, J., \& {Zerilli}, F. 1972, Physical
  Review Letters, 28, 1352

\bibitem[{{Death}(1978)}]{1978PhRvD..18..990D}
{Death}, P.~D. 1978, \prd, 18, 990

\bibitem[{{D'eath} \& {Payne}(1992)}]{1992PhRvD..46..694D}
{D'eath}, P.~D., \& {Payne}, P.~N. 1992, \prd, 46, 694

\bibitem[{{Deffayet} \& {Menou}(2007)}]{2007ApJ...668L.143D}
{Deffayet}, C., \& {Menou}, K. 2007, \apjl, 668, L143

\bibitem[{{Dolan}(2008)}]{2008CQGra..25w5002D}
{Dolan}, S.~R. 2008, Classical and Quantum Gravity, 25, 235002

\bibitem[{{Dolan} \& {Ottewill}(2011)}]{2011PhRvD..84j4002D}
{Dolan}, S.~R., \& {Ottewill}, A.~C. 2011, \prd, 84, 104002

\bibitem[{{Dorband} {et~al.}(2006){Dorband}, {Berti}, {Diener}, {Schnetter}, \&
  {Tiglio}}]{2006PhRvD..74h4028D}
{Dorband}, E.~N., {Berti}, E., {Diener}, P., {Schnetter}, E., \& {Tiglio}, M.
  2006, \prd, 74, 084028

\bibitem[{{Echeverria}(1989)}]{1989PhRvD..40.3194E}
{Echeverria}, F. 1989, \prd, 40, 3194

\bibitem[{{Einstein}(1936)}]{1936Sci....84..506E}
{Einstein}, A. 1936, Science, 84, 506

\bibitem[{{Ferrari} \& {Mashhoon}(1984)}]{1984PhRvD..30..295F}
{Ferrari}, V., \& {Mashhoon}, B. 1984, \prd, 30, 295

\bibitem[{{Field} {et~al.}(2010){Field}, {Hesthaven}, \&
  {Lau}}]{2010PhRvD..81l4030F}
{Field}, S.~E., {Hesthaven}, J.~S., \& {Lau}, S.~R. 2010, \prd, 81, 124030

\bibitem[{{Finn}(1992)}]{1992PhRvD..46.5236F}
{Finn}, L.~S. 1992, \prd, 46, 5236

\bibitem[{{Flanagan} \& {Hughes}(1998)}]{1998PhRvD..57.4535F}
{Flanagan}, {\'E}.~{\'E}., \& {Hughes}, S.~A. 1998, \prd, 57, 4535

\bibitem[{{Fodor} \& {R{\'a}cz}(2003)}]{2003PhRvD..68d4022F}
{Fodor}, G., \& {R{\'a}cz}, I. 2003, \prd, 68, 044022

\bibitem[{{Futterman} {et~al.}(1988){Futterman}, {Handler}, \&
  {Matzner}}]{1988sfbh.book.....F}
{Futterman}, J.~A.~H., {Handler}, F.~A., \& {Matzner}, R.~A. 1988, {Scattering
  from black holes}

\bibitem[{{Gleiser} {et~al.}(2008){Gleiser}, {Price}, \&
  {Pullin}}]{2008CQGra..25g2001G}
{Gleiser}, R.~J., {Price}, R.~H., \& {Pullin}, J. 2008, Classical and Quantum
  Gravity, 25, 072001

\bibitem[{{Goebel}(1972)}]{1972ApJ...172L..95G}
{Goebel}, C.~J. 1972, \apjl, 172, L95

\bibitem[{{Hadar} {et~al.}(2011){Hadar}, {Kol}, {Berti}, \&
  {Cardoso}}]{2011PhRvD..84d7501H}
{Hadar}, S., {Kol}, B., {Berti}, E., \& {Cardoso}, V. 2011, \prd, 84, 047501

\bibitem[{{Holz} \& {Hughes}(2005)}]{2005ApJ...629...15H}
{Holz}, D.~E., \& {Hughes}, S.~A. 2005, \apj, 629, 15

\bibitem[{{Jorstad} {et~al.}(2001){Jorstad}, {Marscher}, {Mattox}, {Wehrle},
  {Bloom}, \& {Yurchenko}}]{2001ApJS..134..181J}
{Jorstad}, S.~G., {Marscher}, A.~P., {Mattox}, J.~R., {Wehrle}, A.~E., {Bloom},
  S.~D., \& {Yurchenko}, A.~V. 2001, \apjs, 134, 181

\bibitem[{{Jorstad} {et~al.}(2005){Jorstad}, {Marscher}, {Lister}, {Stirling},
  {Cawthorne}, {Gear}, {G{\'o}mez}, {Stevens}, {Smith}, {Forster}, \&
  {Robson}}]{2005AJ....130.1418J}
{Jorstad}, S.~G., {et~al.} 2005, \aj, 130, 1418

\bibitem[{{Kaloper}(2005)}]{2005PhRvD..71h6003K}
{Kaloper}, N. 2005, \prd, 71, 086003

\bibitem[{{Karas} \& {{\v S}ubr}(2001)}]{2001A&A...376..686K}
{Karas}, V., \& {{\v S}ubr}, L. 2001, \aap, 376, 686

\bibitem[{{Keshet} \& {Ben-Meir}(2012)}]{2012arXiv1207.2460K}
{Keshet}, U., \& {Ben-Meir}, A. 2012, ArXiv e-prints, arxiv:1207.2460

\bibitem[{{Kocsis} {et~al.}(2006){Kocsis}, {Frei}, {Haiman}, \&
  {Menou}}]{2006ApJ...637...27K}
{Kocsis}, B., {Frei}, Z., {Haiman}, Z., \& {Menou}, K. 2006, \apj, 637, 27

\bibitem[{{Kocsis} {et~al.}(2008){Kocsis}, {Haiman}, \&
  {Menou}}]{2008ApJ...684..870K}
{Kocsis}, B., {Haiman}, Z., \& {Menou}, K. 2008, \apj, 684, 870

\bibitem[{{Kocsis} \& {Levin}(2012)}]{2012PhRvD..85l3005K}
{Kocsis}, B., \& {Levin}, J. 2012, \prd, 85, 123005

\bibitem[{{Kovacs} \& {Thorne}(1978)}]{1978ApJ...224...62K}
{Kovacs}, Jr., S.~J., \& {Thorne}, K.~S. 1978, \apj, 224, 62

\bibitem[{{Leaver}(1986)}]{1986PhRvD..34..384L}
{Leaver}, E.~W. 1986, \prd, 34, 384

\bibitem[{{LIGO Scientific Collaboration} {et~al.}(2012){LIGO Scientific
  Collaboration}, {Virgo Collaboration}, {Abadie}, {Abbott}, {Abbott},
  {Abbott}, {Abernathy}, {Accadia}, {Acernese}, {Adams}, \&
  et~al.}]{2012A&A...539A.124L}
{LIGO Scientific Collaboration} {et~al.} 2012, \aap, 539, A124

\bibitem[{{Lousto} \& {Zlochower}(2011)}]{2011PhRvL.106d1101L}
{Lousto}, C.~O., \& {Zlochower}, Y. 2011, \prl, 106, 041101

\bibitem[{{MacLeod} \& {Hogan}(2008)}]{2008PhRvD..77d3512M}
{MacLeod}, C.~L., \& {Hogan}, C.~J. 2008, \prd, 77, 043512

\bibitem[{{Mao} \& {Paczynski}(1991)}]{1991ApJ...374L..37M}
{Mao}, S., \& {Paczynski}, B. 1991, \apjl, 374, L37

\bibitem[{{Matzner} {et~al.}(1985){Matzner}, {Dewitte-Morette}, {Nelson}, \&
  {Zhang}}]{1985PhRvD..31.1869M}
{Matzner}, R.~A., {Dewitte-Morette}, C., {Nelson}, B., \& {Zhang}, T.-R. 1985,
  \prd, 31, 1869

\bibitem[{{McKinney} \& {Blandford}(2009)}]{2009MNRAS.394L.126M}
{McKinney}, J.~C., \& {Blandford}, R.~D. 2009, \mnras, 394, L126

\bibitem[{{Milotti}(2002)}]{2002physics...4033M}
{Milotti}, E. 2002, ArXiv Physics e-prints, physics/0204033

\bibitem[{{Nakamura}(1998)}]{1998PhRvL..80.1138N}
{Nakamura}, T.~T. 1998, Physical Review Letters, 80, 1138

\bibitem[{{Naoz} {et~al.}(2012){Naoz}, {Kocsis}, {Loeb}, \&
  {Yunes}}]{2012arXiv1206.4316N}
{Naoz}, S., {Kocsis}, B., {Loeb}, A., \& {Yunes}, N. 2012, ArXiv e-prints,
  arxiv:1206.4316

\bibitem[{{Nissanke} {et~al.}(2010){Nissanke}, {Holz}, {Hughes}, {Dalal}, \&
  {Sievers}}]{2010ApJ...725..496N}
{Nissanke}, S., {Holz}, D.~E., {Hughes}, S.~A., {Dalal}, N., \& {Sievers},
  J.~L. 2010, \apj, 725, 496

\bibitem[{{Nissanke} {et~al.}(2012){Nissanke}, {Kasliwal}, \&
  {Georgieva}}]{2012arXiv1210.6362N}
{Nissanke}, S., {Kasliwal}, M., \& {Georgieva}, A. 2012, ArXiv e-prints,
  arXiv:1210.6362

\bibitem[{{O'Leary} {et~al.}(2009){O'Leary}, {Kocsis}, \&
  {Loeb}}]{2009MNRAS.395.2127O}
{O'Leary}, R.~M., {Kocsis}, B., \& {Loeb}, A. 2009, \mnras, 395, 2127

\bibitem[{{Press}(1971)}]{1971ApJ...170L.105P}
{Press}, W.~H. 1971, \apjl, 170, L105

\bibitem[{{Press}(1978)}]{1978ComAp...7..103P}
---. 1978, Comments on Astrophysics, 7, 103

\bibitem[{{Press} {et~al.}(1992){Press}, {Teukolsky}, {Vetterling}, \&
  {Flannery}}]{1992nrfa.book.....P}
{Press}, W.~H., {Teukolsky}, S.~A., {Vetterling}, W.~T., \& {Flannery}, B.~P.
  1992, {Numerical recipes in FORTRAN. The art of scientific computing}

\bibitem[{{Price}(1972)}]{1972PhRvD...5.2419P}
{Price}, R.~H. 1972, \prd, 5, 2419

\bibitem[{{Pulido} \& {Caranti}(2000)}]{2000JAtS...57.1473P}
{Pulido}, M., \& {Caranti}, G. 2000, J. Atmos. Sci., 57, 1473

\bibitem[{{Santamar{\'{\i}}a~et.~al.}(2010)}]{2010PhRvD..82f4016S}
{Santamar{\'{\i}}a~et.~al.}, L. 2010, \prd, 82, 064016

\bibitem[{{Schnittman}(2010)}]{2010ApJ...724...39S}
{Schnittman}, J.~D. 2010, \apj, 724, 39

\bibitem[{{Schutz}(1986)}]{1986Natur.323..310S}
{Schutz}, B.~F. 1986, \nat, 323, 310

\bibitem[{{Segalis} \& {Ori}(2001)}]{2001PhRvD..64f4018S}
{Segalis}, E.~B., \& {Ori}, A. 2001, \prd, 64, 064018

\bibitem[{{Seto} \& {Muto}(2010)}]{2010PhRvD..81j3004S}
{Seto}, N., \& {Muto}, T. 2010, \prd, 81, 103004

\bibitem[{{Seto} \& {Muto}(2011)}]{2011MNRAS.415.3824S}
---. 2011, \mnras, 415, 3824

\bibitem[{{Sikora} {et~al.}(2005){Sikora}, {Begelman}, {Madejski}, \&
  {Lasota}}]{2005ApJ...625...72S}
{Sikora}, M., {Begelman}, M.~C., {Madejski}, G.~M., \& {Lasota}, J. 2005, \apj,
  625, 72

\bibitem[{{Urry} \& {Padovani}(1995)}]{1995PASP..107..803U}
{Urry}, C.~M., \& {Padovani}, P. 1995, \pasp, 107, 803

\bibitem[{{{\v S}ubr} \& {Karas}(1999)}]{1999A&A...352..452S}
{{\v S}ubr}, L., \& {Karas}, V. 1999, \aap, 352, 452

\bibitem[{{Vaidya}(1951)}]{1951PhRv...83...10V}
{Vaidya}, P.~C. 1951, Phys. Rev., 83, 10

\bibitem[{{van Meter} {et~al.}(2010){van Meter}, {Wise}, {Miller}, {Reynolds},
  {Centrella}, {Baker}, {Boggs}, {Kelly}, \&
  {McWilliams}}]{2010ApJ...711L..89V}
{van Meter}, J.~R., {et~al.} 2010, \apjl, 711, L89

\bibitem[{{Virbhadra} \& {Ellis}(2000)}]{2000PhRvD..62h4003V}
{Virbhadra}, K.~S., \& {Ellis}, G.~F.~R. 2000, \prd, 62, 084003

\bibitem[{{Watson}(1918)}]{watson}
{Watson}, G.~N. 1918, Proc. London Math. Soc., 2, 116

\bibitem[{{Wen}(2003)}]{2003ApJ...598..419W}
{Wen}, L. 2003, \apj, 598, 419

\bibitem[{{Yang} {et~al.}(2012){Yang}, {Nichols}, {Zhang}, {Zimmerman},
  {Zhang}, \& {Chen}}]{2012arXiv1207.4253Y}
{Yang}, H., {Nichols}, D.~A., {Zhang}, F., {Zimmerman}, A., {Zhang}, Z., \&
  {Chen}, Y. 2012, \prd, 86, 104006

\bibitem[{{Zengino{\u g}lu} \& {Galley}(2012)}]{2012PhRvD..86f4030Z}
{Zengino{\u g}lu}, A., \& {Galley}, C.~R. 2012, \prd, 86, 064030

\end{thebibliography}

\end{document}